 \documentstyle[12pt]{article}

\catcode`\@=11
\long\def\@makefntext#1{
\protect\noindent \hbox to 3.2pt {\hskip-.9pt
$^{{\@thefnmark}}$\hfil}#1\hfill}               

\def\@makefnmark{\hbox to 0pt{$^{\@thefnmark}$\hss}}    

\def\ps@myheadings{\let\@mkboth\@gobbletwo
\def\@oddhead{\hbox{}
\rightmark\hfil\thepage}
\def\@oddfoot{}\def\@evenhead{\thepage\hfil
\leftmark\hbox{}}\def\@evenfoot{}
\def\sectionmark##1{}\def\subsectionmark##1{}}

\renewenvironment{thebibliography}[1]
        {\frenchspacing
         \baselineskip=18pt
         \begin{list}{\arabic{enumi}.}
        {\usecounter{enumi}\setlength{\parsep}{0pt}
         \setlength{\leftmargin 12.7pt}{\rightmargin 0pt} 
         \setlength{\itemsep}{0pt} \settowidth
        {\labelwidth}{#1.}\sloppy}}{\end{list}}

\newcommand{\tcaption}[1]{
        \refstepcounter{table}
        \setbox\@tempboxa = \hbox{\footnotesize Table~\thetable. #1}
        \ifdim \wd\@tempboxa > 5in
           {\begin{center}
        \parbox{5in}{\footnotesize\smalllineskip Table~\thetable. #1}
            \end{center}}
        \else
             {\begin{center}
             {\footnotesize Table~\thetable. #1}
              \end{center}}
        \fi}

\def\@citex[#1]#2{\if@filesw\immediate\write\@auxout
        {\string\citation{#2}}\fi
\def\@citea{}\@cite{\@for\@citeb:=#2\do
        {\@citea\def\@citea{,}\@ifundefined
        {b@\@citeb}{{\bf ?}\@warning
        {Citation `\@citeb' on page \thepage \space undefined}}
        {\csname b@\@citeb\endcsname}}}{#1}}

\newif\if@cghi
\def\cite{\@cghitrue\@ifnextchar [{\@tempswatrue
        \@citex}{\@tempswafalse\@citex[]}}
\def\citelow{\@cghifalse\@ifnextchar [{\@tempswatrue
        \@citex}{\@tempswafalse\@citex[]}}
\def\@cite#1#2{{$\null^{#1}$\if@tempswa\typeout
        {IJCGA warning: optional citation argument
        ignored: `#2'} \fi}}

\def\pmb#1{\setbox0=\hbox{#1}
        \kern-.025em\copy0\kern-\wd0
        \kern.05em\copy0\kern-\wd0
        \kern-.025em\raise.0433em\box0}

\def\anti{\buildrel \wedge \over }
\def\symm{\buildrel \sim \over }
\def\ps{{}^p\!}
\def\vect{\buildrel \rightharpoonup \over }
\def\tens{{}^s\!}

\def\a{\alpha }
\def\be{\beta }
\def\g{\gamma }
\def\d{{\cal D}}
\def\de{\delta }
\def\Dep{\varpi }
\def\e{\epsilon }
\def\f{{\cal P}}
\def\fc{{\xi_\bot}}
\def\h{{\cal H}}
\def\ve{\varepsilon }
\def\th{e}

\def\k{\kappa }
\def\la{\lambda }
\def\La{\Lambda }

\def\lan{\langle }
\def\na{\nabla \! }

\def\vph{\varphi }
\def\G{{\mit\Gamma}}
\def\p{\partial }
\def\ol{\overline }
\def\ran{\rangle }
\def\s{\sigma }
\def\tc{\zeta }

\def\ul{\underline }

\def\ri{\uppercase\expandafter{\romannumeral1 }}
\def\rj{\uppercase\expandafter{\romannumeral2 }}
\def\rk{\uppercase\expandafter{\romannumeral3 }}
\def\rl{\uppercase\expandafter{\romannumeral4 }}
\def\rx{\uppercase\expandafter{\romannumeral5 }}
\def\ry{\uppercase\expandafter{\romannumeral6 }}
\def\Ri{\romannumeral1 }
\def\Rj{\romannumeral2 }

\renewcommand{\theequation}{\arabic{section}.\arabic{equation}}

\title{\Large\bf Hamiltonian analysis of Poincar\'e
gauge theory scalar modes}
\author{\large
Hwei-Jang Yo\thanks{Electronic address: s2234003@twncu865.ncu.edu.tw}\ \
and James M. Nester\thanks{Electronic address: nester@joule.phy.ncu.edu.tw} \\
\large\it Department of Physics and Center for Complex Systems,\\
\large\it National Central University, Chungli 320, Taiwan, ROC Fy; 04.50 +h\hspace*{\fill}\\
\rm\normalsize
Short title: {\bf Hamiltonian analysis of PGT scalar modes}\hspace*{\fill}}
\date{\today}

\begin{document}
\maketitle
 \begin{abstract}
The Hamiltonian constraint formalism is used to obtain the first
explicit complete analysis of non-trivial viable dynamic modes for the
Poincar\'e gauge theory of gravity.  Two modes with propagating
spin-zero torsion are analyzed.  The explicit form of the Hamiltonian
is presented.  All constraints are obtained and classified.  The
Lagrange multipliers are derived.  It is shown that a massive
spin-$0^-$ mode has normal dynamical propagation but the associated
massless $0^-$ is pure gauge.  The spin-$0^+$ mode investigated here
is also viable in general. Both modes exhibit a simple type of ``constraint
bifurcation'' for certain special field/parameter values.
\end{abstract}

\section{Introduction}
 \setcounter{equation}{0}

The Poincar\'e gauge theory of gravity (PGT)\cite{Hehl80,Hayash80},
based on a Riemann-Cartan geometry, allows for dynamic torsion in
addition to curvature.  Because of its gauge structure and geometric
properties it was regarded as an attractive alternative to general
relativity (GR).  It was quickly realized that the theory had physical
difficulties with generic values for its ten coupling parameters (see,
e.g., references\cite{Hayash80,Sezgin80a,Sezgin80b,Battit85,Chern92}).
Consequently investigators have looked for restrictions giving viable
sets of PGT parameters.  By utilizing certain theoretical tests (e.g.,
``no-ghosts'' or ``no-tachyons''), sets of constraints on the
parameters and possible viable PGT modes were obtained (see, e.g.,
references\cite{Hayash80,Sezgin80a,Sezgin80b,Battit85,Kuhfus86}).  These
investigations naturally used the weak-field approximation and
linearization of the theory, avoiding the inherently highly nonlinear
complications in the PGT.

Subsequently, investigations\cite{Dimaki89} indicated that certain
degeneracies of the Hessian matrix, thought to be necessary for a
viable theory, appeared to lead to difficulties with the initial value
problem.  Then a shock-wave analysis\cite{Lemke90,Hecht90} concluded
that these same degeneracies allowed tachyonic propagating modes.
These difficulties were more recently reconsidered\cite{HNZ94}.  It was found
that, when all the constraints are taken into account, there are no
problems --- at the linear order.

However, the PGT Hamiltonian analysis tells a more complicated story.
The virtue of the Hamiltonian analysis is that it provides a clear
vision of the possible degeneracies, corresponding constraints, and
true degrees of freedom of a theory.  Its application to the nonlinear
PGT is revealing.  It shows that the nonlinear behavior of the PGT can
be --- and, through a phenomenon referred to as ``constraint
bifurcation''\cite{Cheng88}, very likely will be --- qualitatively different
from the linearized one in the number and type of constraints, so the
linearized ``good modes'' may very well not be viable in the full
nonlinear theory\cite{Chen98}.  Hence, in order to understand the
subtle behavior of the PGT and search for modes which truly have good
propagation, it is important that the analysis consider the full
nonlinear scope of the theory.

It has long been known that non-linear couplings involving higher
spins are problematical\cite{VZ69,AD71}.  However, spin-zero modes were
expected to be problem free.  For the PGT, it has subsequently been
verified that two four-parameter subclasses with only spin-zero
propagating modes really do have a well posed initial value problem
without any tachyonic propagation\cite{Hecht96}.

That leads us to examine the Hamiltonian formalism for
some representative dynamic spin-zero PGT modes in order to see how (and
indeed whether) the modes really manage to avoid the nearly ubiquitous
(and almost certainly fatal) nonlinear ``constraint bifurcation''.

Such an investigation seems to be an essential prelude to a search for
any higher spin nonlinear-problem-free modes in the PGT.  Our results
presented here gives, to our knowledge, the only complete Hamiltonian
analysis of PGT modes believed to be (linearly and nonlinearly)
viable.

Also, it may be worthwhile to mention, certain interesting results
which were found in studies of higher-derivative gravity (see, e.g.,
references\cite{Stelle78,Teyssa83,Whitt84,Hindaw96}).
These theories contain a non-ghost scalar field, in addition to the usual
graviton of GR.  Some pure gravity inflationary models for the
Universe were proposed based on such theories, although the issue
still remains controversial\cite{Starob80,Mijic86,Barrow88,Simon92}.  This has
provided additional
motivation for looking for similar situations in the PGT, but it must
be kept in mind that the principle and structure between the PGT and
higher-derivative gravity are quite different.

The paper is organized as follows.  In section 2 we review the basic
elements of the PGT and introduce its Lagrangian and field equations.
In section 3 we use the Dirac theory for constrained Hamiltonian
systems in the excellent ``if'' constraint formulation developed by
Blagojevi\'c and Nikoli\'c\cite{Blagoj83,Nikoli84}.  The primary
constraints, including ten ``sure'' primary constraints and thirty
so-called ``primary if-constraints'', are found.  The total
Hamiltonian density, including the canonical Hamiltonian density
and all possible primary constraints, is derived.  In section 4 we
consider two very degenerate spin-zero modes, each having only one of
the six parameters of the quadratic curvature parts being non-zero.
In particular, we consider a spin-$0^+$, a massive spin-$0^-$ and the
associated massless spin-$0^-$ modes.  The Lagrange multipliers are
derived.  It is shown that the massless spin-$0^-$ mode is unphysical,
being pure gauge.  In section 5 the degrees of freedom are counted.
The multipliers are shown to be exactly the missing `velocities'.
Their effects in the spin-zero modes are discussed.  In both
scalar modes we find a simple
type of ``constraint bifurcation'' phenomenon, which is connected with changes
in the nature of the constraint reduced from
the Lorentz rotation parts of the canonical Hamiltonian density.
The similarity between the spin-$0^+$ case and
higher-derivative gravity is noted.  In the final section we present
our conclusions.

Throughout the paper our PGT conventions are basically the same as
Hehl's\cite{Hehl80}.  We have made a few adjustments to accommodate
the translation of the Hamiltonian ``if'' constraint formalism to
these conventions.  The latin indices are coordinate (holonomic)
indices, whereas the greek indices are orthonormal frame (anholonomic)
indices.  The first letters of both alphabets ($a, b, c, \dots$; $\a,
\be, \g, \dots$) run over 1, 2, 3, whereas the later ones run over 0,
1, 2, 3.  Furthermore, $\eta _{\mu \nu}=$diag($-,+,+,+$); $\e ^{\mu
\nu \g \de}$ is the completely antisymmetric tensor with $\e ^{{\hat
0}{\hat 1}{\hat 2}{\hat 3}}=-1$.  The meaning of a bar over a greek
index is adopted from Blagojecvi\'c and Nickoli\'c \cite{Nikoli84}.


\section{Poincar\'e gauge theory of gravitation}
\setcounter{equation}{0}

 In the PGT there are two sets of gauge potentials, the orthonormal
 frame field (tetrads) $\th _i{}^\mu$ and the metric-compatible connection
 $\G _{i\mu}{}^\nu$, which are associated with the translation and the
Lorentz
 subgroups of the Poincar\'e gauge group, respectively. The associated
 field strengths are the torsion
 \begin{equation}
  T_{ij}{}^\mu =2(\p _{[i}\th _{j]}{}^\mu
  +\G _{[i|\nu }{}^\mu \th _{|j]}{}^\nu),
 \end{equation}
 and the curvature
 \begin{equation}
  R_{ij\mu}{}^\nu =2(\p _{[i}\G _{j]\mu}{}^\nu
  +\G _{[i|\s }{}^\nu \G _{|j]\mu}{}^\s ),
 \end{equation}
 which satisfy the  Bianchi identities
 \begin{eqnarray}
  &&\na _{[i}T_{jk]}{}^\mu \equiv R_{[ijk]}{}^\mu,\\
  &&\na _{[i}R_{jk]}{}^{\mu \nu}\equiv 0.
 \end{eqnarray}
 The conventional form of the action, which is invariant under the
 Poincar\'e gauge group, has the form
 \begin{equation}
  A=\int d^4x\th(L_M+L_G),
 \end{equation}
 where $L_M$ stands for the matter Lagrangian density (which
determines the energy-momentum and spin source currents),
$L_G$ denotes
the gravitational Lagrangian density, and $\th =det(\th _i{}^\mu )$.
In this paper we are concerned with the gravitational propagating
modes,  hence we omit
the matter Lagrangian density, so $L_G$ is considered
as the source-free total Lagrangian. Varying
with respect to the potentials then gives the (vacuum) field equations,
 \begin{eqnarray}
  &&\na _j\f_\mu {}^{ij}-\ve _\mu {}^i=0,\\
  &&\na _j\f_{\mu \nu}{}^{ij}-\ve _{\mu \nu}{}^i=0,
 \end{eqnarray}
 with the field momenta
 \begin{eqnarray}
  \f _\mu {}^{ij}&:=&{\p \th L_G\over \p \p _j\th _i{}^\mu}
  =2{\p \th L_G\over \p T_{ji}{}^\mu},\\
  \f _{\mu \nu}{}^{ij}&:=&{\p \th L_G\over \p \p _j\G _i{}^{\mu \nu}}
  =2{\p \th L_G\over \p R_{ji}{}^{\mu \nu}},
 \end{eqnarray}
 and
 \begin{eqnarray}
  \ve _\mu {}^i&:=&e^i{}_\mu \th L_G-T_{\mu j}{}^\nu \f_\nu {}^{ji}
  -R_{\mu j}{}^{\nu \s}\f _{\nu \s}{}^{ji},\\
  \ve _{\mu \nu}{}^i&:=&\f _{[\nu \mu]}{}^i.
 \end{eqnarray}
The Lagrangian is chosen (as usual) to
be at most of quadratic order in the field strengths,
 then the field momenta are
linear in the field strengths:
 \begin{eqnarray}
  \f _\mu {}^{ij}&=&{\th \over l^2}\sum^3_{k=1} a_k
  {\buildrel {(k)} \over T}{}^{ji}{}_\mu,\\
  \f _{\mu \nu}{}^{ij}&=&-{a_0\th \over l^2}e^i{}_{[\mu}e^j{}_{\nu ]}
  +{\th \over \k}\sum^6_{k=1}b_k
  {\buildrel {(k)} \over R}{}^{ji}{}_{\mu \nu},
 \end{eqnarray}
the three $\displaystyle {{\buildrel {(k)} \over T}{}^{ji}{}_\mu}$
and the six $\displaystyle { {\buildrel {(k)} \over R}{}^{ji}{}_{\mu
\nu}}$ are the algebraically irreducible parts of the torsion and the
curvature, respectively. The reciprocal frames $e^i{}_\mu$ and
$\th _i{}^\mu$ satisfy $e^i{}_\mu \th _i{}^\nu =\de _\mu {}^\nu$
and $e^i{}_\mu \th _j{}^\mu =\de _j{}^i$; the coordinate metric
is defined by $g_{ij}=\th _i{}^\mu \th _j{}^\nu \eta _{\mu \nu}$.
The $a_k$ and $b_k$ are free coupling parameters. Due to
the Bach-Lanczos identity only five of the six $b_k$'s are
independent.  $a_0$ is the coupling parameter of the scalar curvature
$R:=R_{\mu \nu}{}^{\nu \mu}$. For the Hamiltonian
formulation we associate the {\it canonical momenta} with certain components
of the covariant field momenta:
 \begin{eqnarray}
  \pi ^i{}_\mu &\equiv &\f _\mu {}^{i0},\label{tm} \\
  \pi ^i{}_{\mu \nu}&\equiv &\f _{\mu \nu}{}^{i0}.\label{cm}
 \end{eqnarray}


\section{Primary constraints and total Hamiltonian}
\setcounter{equation}{0}
In this section we present the primary constraints and the total
Hamiltonian density of the PGT in terms of the decomposition of the canonical
variables and fields.  First of all, one obtains the ``sure'' primary
constraints
 \begin{eqnarray}
  \pi ^0{}_\mu &\approx &0,\label{stpc} \\
  \pi ^0{}_{\mu \nu}&\approx &0.\label{scpc}
 \end{eqnarray}
These constraints just reflect the fact that the torsion and the curvature are
defined as the antisymmetric derivatives of $\th _i{}^\mu$ and $\G
_i{}^{\mu \nu}$; they do not involve the ``velocities'' ${\dot
\th}_0{}^\mu$
and ${\dot \G}_0{}^{\mu \nu}$. One will obtain further primary constraints
if the Lagrangian density is singular with respect to the remaining
``velocities'', $\dot\th _a{}^\mu$, and $\dot\G _a{}^{\mu \nu}$. (Such
constraints, so-called ``primary if-constraints'', result
from certain vanishing coupling parameter combinations.) The total
Hamiltonian density is of the form
 \begin{equation}
  \h _{\hbox{\footnotesize tot}}=\h _{\hbox{\footnotesize can}}
  +u_0{}^\mu \pi ^0{}_\mu
  +{1\over 2}u_0{}^{\mu \nu}\pi ^0{}_{\mu \nu}+u^A\phi _A,
  \label{thd}
 \end{equation}
where the $\phi _A$ are
the primary ``if'' constraints and the $u$'s
denote the associated {\it Lagrange multipliers}; $\h
_{\hbox{\footnotesize can}}$ stands for the canonical
Hamiltonian density which will be specified below.

Before we proceed to obtain the explicit form of the canonical
Hamiltonian density and $\phi _A$,
it is convenient to define the decomposition of related variables and
functions. We essentially follow the techniques developed by
Blagojevi\'c and Nikoli\'c\cite{Blagoj83}.
 Let us note that the components of the unit normal {\bf n}
to the $x^0=$constant hypersurface, with respect to the orthonormal
frame, are given by
 \begin{equation}
  n_\mu :={-e^0{}_\mu \over \sqrt {-g^{00}}}.
 \end{equation}
A vector, e.g., $V_\mu$, can be decomposed naturally into the
orthogonal and parallel components with respect to the orthonormal
frame indices:
 \begin{eqnarray}
  V_\mu &=&-V_\bot n_\mu +V_{\ol \mu },\\
  V_\bot &\equiv &V_\mu n^\mu,\\
  V_{\ol {\mu}}&\equiv &V_\nu(\de _\mu {}^\nu +n_\mu n^\nu ).
 \end{eqnarray}
 One can easily extend the decomposition to any tensors with
 orthonormal frame indices. The lapse and shift functions can be
 written as
 \begin{eqnarray}
  N&\equiv &{1\over \sqrt {-g^{00}}}=-n_\mu \th _0{}^\mu ,\\
  N^a&\equiv &-{g^{0a}\over g^{00}}=\th _0{}^\mu e_{\ol \mu }{}^a,
 \end{eqnarray}
 and $\th =NJ$,  
 where $J$ is the determinant of the 3-metric.
 Defining the convenient ``parallel'' canonical momenta
  \begin{equation}
   \pi ^{\ol \s}{}_\mu \equiv \th _a{}^\s \pi ^a{}_\mu,\quad
   \pi ^{\ol \s}{}_{\mu \nu}\equiv \th _a{}^\s \pi ^a
   {}_{\mu \nu},
  \end{equation}
 which satisfy $\pi ^{\ol \s}{}_\mu n_\s =0$, $\pi ^{\ol \s}{}_{\mu
 \nu}n_\s =0$, the canonical Hamiltonian density,
  \begin{equation}
   \h _{\hbox{\footnotesize can}}
   =\pi ^a{}_\mu {\dot \th}_a{}^\mu +{1\over 2}\pi ^a{}
   _{\mu \nu}{\dot \G}_a{}^{\mu \nu}-\th L,
  \end{equation}
 can be rewritten in the so-called Dirac-ADM
 form\cite{Dirac64,Hanson76,Sunder82},
  \begin{equation}
   \h _{\hbox{\footnotesize can}}=N\h _\bot +N^a\h _a
    +{1\over 2}\G_0{}^{\mu \nu}\h _{\mu \nu}+\p _a\d ^a,\label{chd}
  \end{equation}
 which is linear in $N$ and $N^a$. The other quantities are given
 by
  \begin{eqnarray}
           \h _\bot =&&\pi ^{\ol \s}{}_\mu T_{\bot \ol \s}{}^
           \mu +{1\over 2}\pi ^{\ol \s}{}_{\mu \nu}R_{\bot \ol \s}
           {}^{\mu \nu}-JL-n^\mu \na _a\pi ^a{}_\mu, \\
           \h _a=&&\pi ^b{}_\mu T_{ab}{}^\mu +{1\over 2}\pi ^b{}_
           {\mu \nu}R_{ab}{}^{\mu \nu}-\th _a{}^\mu \na _b
           \pi ^b{}_\mu, \\
           \h _{\mu \nu}=&&\pi ^a{}_\mu \th _{a\nu}-\pi ^a{}_\nu
           \th _{a\mu}-\na _a\pi ^a{}_{\mu \nu},\\
           \d ^a=&&\pi ^a{}_\mu \th _0{}^\mu +{1\over 2}\pi
           ^a{}_{\mu \nu }\G _0{}^{\mu \nu }.
  \end{eqnarray}
  Only the super-Hamiltonian $\h _\bot$ is involved in dynamical
  evolution, the super-momenta $\h _a$ and Lorentz rotation parts
  $\h _{\mu \nu}$ are kinematical generators, consequently  we
concentrate  on $\h _\bot$ when consistency conditions are calculated.
  By utilizing the forms of the total and canonical Hamiltonian density
 (\ref{thd})
 and (\ref{chd}) in the consistency conditions for the primary constraints
  (\ref{stpc}) and (\ref{scpc}) we  obtain the {\it secondary constraints} (SC)
 \begin{equation}
   \h _\bot \approx 0,\quad \h _a \approx 0,\quad
   \h _{\mu \nu}\approx 0.
 \end{equation}
Since it is easy to check that the primary constraints (\ref{stpc}) and
(\ref{scpc}) are first-class, i.e., $\pi ^0{}_\mu$ and $\pi ^0{}_{\mu \nu}$ are
unphysical variables, by using the Hamilton equation of motion we can infer
that the multipliers $u_0{}^\mu$ and $u_0{}^{\mu \nu}$ are indeed equal to $\dot
\th _0{}^\mu$ and $\dot \G _0{}^{\mu \nu}$. They are dynamically undetermined
pure gauge multipliers.

 It is necessary to understand the relation between the
 canonical momenta and the velocities well before
 one can make $\h _\bot$ more apparent. Let us consider the torsion
 momenta $\pi _{\ol \mu \nu}$  at first. $\pi _{\ol \mu \nu}$ can be
 decomposed into four irreducible parts as follows:
 \begin{eqnarray}
  \pi _{\ol \mu \nu}=&&-n_\nu \pi _{\ol \mu \bot}+\pi _{\ol {\mu \nu}}
                      \nonumber \\
                    =&&-n_\nu \pi _{\ol \mu \bot}
                    +{\anti \pi}_{\ol {\mu \nu}}
                    +{\symm \pi}_{\ol {\mu \nu}}+{1\over 3}
  \eta _{\ol {\mu \nu}}\pi,
 \end{eqnarray}
where ${\anti \pi}_{\ol {\mu \nu}}$, $\pi$ and ${\symm \pi}_{\ol {\mu
\nu}}$ are the antisymmetric part, trace part and symmetric-traceless
part of $\pi _{\ol {\mu \nu}}$, respectively.  Manipulating the
definition of the torsion momenta (\ref{tm}), one finds the following
relations between the different parts of the canonical momenta and the
corresponding parts of the velocities $T_{\bot \ol \mu \nu}$:
{\setcounter{enumi}{\value{equation}}
 \addtocounter{enumi}{1}
 \setcounter{equation}{0}
 \renewcommand{\theequation}{\arabic{section}.\theenumi\alph{equation}}
 \begin{eqnarray}
  \phi _{\ol \mu \bot }
  \equiv &&{\pi _{\ol \mu \bot }\over J}-
           {1\over 3l^2}(a_1-a_2){\vect T}_{\ol \mu}
          =-{1\over 3l^2}(2a_1+a_2)
              T_{\bot \ol \mu \bot },\\
  {\anti \phi }_{\ol {\mu \nu }}\equiv &&{{\anti \pi }
           _{\ol {\mu \nu }}\over J}-{1\over 3l^2}(a_1-a_3)T_{
           \ol {\mu \nu }\bot }
  =-{1\over 3l^2}(a_1+2a_3)
           T_{\bot [\ol {\mu \nu}]},\\
  {\symm \phi }_{\ol {\mu \nu }}
  \equiv &&{{\symm \pi}_{\ol {\mu \nu}}\over J}=-{a_1\over l^2}
           T_{\bot \lan \ol {\mu \nu}\ran},\\
  \phi \equiv &&{\pi \over J}=-{a_2\over l^2}T_{\bot \ol \s}{}^{\ol\s},
 \end{eqnarray}
 \setcounter{equation}{\value{enumi}}}
 where ${\vect T}_{\ol \mu}\equiv T_{\ol{\mu \nu}}{}^{\ol \nu}$,
 and a tensor with two indices contained in the bracket $\lan \, \ran$
 denotes that the tensor is symmetric-traceless with respect to the
two  indices.
 If the parameters take on the critical values: $2a_1+a_2=0$, $a_1+2a_3=0$,
 $a_1=0$, and/or $a_2=0$, one obtains the following ``{\it primary
 if-constraints}'' (PIC): $\phi _{\ol \mu \bot}\approx 0$,
 ${\anti \phi}_{\ol{\mu \nu}}\approx 0$,
 ${\symm \phi}_{\ol{\mu \nu}}\approx 0$, and/or
 $\phi \approx 0$, respectively.

 The curvature momenta $\pi _{\ol \s \mu \nu}$ can be decomposed into
 six irreducible parts:
 \begin{equation}
  \pi _{\ol \s \mu \nu}=\pi _{\ol{\s \mu \nu}}+2\pi _{\ol \s \bot [\ol
  \mu}n_{\nu]},
 \end{equation}
 and
 \begin{eqnarray}
  \pi _{\ol{\mu \nu}\bot}=&&{\anti \pi}_{\ol{\mu \nu}\bot}
                    +{\symm \pi}_{\ol{\mu \nu}\bot}+{1\over 3}
  \eta _{\ol {\mu \nu}}\pi _\bot,\\
  \pi _{\ol{\s \mu \nu}}=&&-{1\over 6}\e _
  {\s \mu \nu \bot}\ps \pi +{\vect \pi}_{[\ol \mu}\eta
  _{\ol \nu ]\ol \s}+{4\over 3}{\tens \pi}_
  {\ol \s [\ol{\mu \nu}]},
 \end{eqnarray}
 where the notation is as follows:
 For a spatial tensor $X_{\ol{\s \mu \nu}}=X_{\ol \s [\ol{\mu \nu}]}$,
 $\ps X\equiv \e ^{\s \mu \nu \bot}X_{\ol{\s \mu \nu}}$,
 ${\vect X}_{\ol \mu}\equiv X_{\ol{\nu \mu}}{}^{\ol \nu}$, and
 ${\tens X}_{\ol{\s \mu \nu}}\equiv X_{(\ol \s |\ol \mu |\ol \nu)}
 -{1\over 2}{\vect X}_{\ol \mu}\eta _{\ol{\nu \s}}
 +{1\over 2}\eta _{\ol \mu (\ol \nu}{\vect X}_{\ol \s )}$ are the
 pseudoscalar part, vector part, and traceless tensor part of
 $X_{\ol{\s \mu \nu}}$, respectively. Identifying the irreducible
 parts of the curvature momenta (\ref{cm}), one finds
{\setcounter{enumi}{\value{equation}}
 \addtocounter{enumi}{1}
 \setcounter{equation}{0}
 \renewcommand{\theequation}{\arabic{section}.\theenumi\alph{equation}}
 \begin{eqnarray}
  \ps \phi \equiv &&{\ps \pi \over J}+{1\over \k}(b_2-b_3)
           {\ps R}_{\circ \bot}
           =-{1\over \k}(b_2+b_3){\ps R}_{\bot \circ},\label{323a}\\
  {\vect \phi}_{\ol \mu}\equiv &&{{\vect \pi}_{\ol \mu}\over J}
           -{1\over \k}(b_4-b_5)R_{\ol \mu \bot}
           ={1\over \k}(b_4+b_5)R_{\bot \ol \mu},\\
  {\tens \phi}_{\ol{\s \mu \nu}}\equiv &&{{\tens \pi}
           _{\ol{\s \mu \nu}}\over J}-{1\over \k}
           (b_1-b_2){\tens R}_{\ol{\mu \nu \s}\bot}
           =-{1\over \k}(b_1+b_2){\tens R}_{\bot \ol{\s \mu \nu}},\\
  \phi _\bot \equiv &&{\pi _\bot \over J}+{3a_0\over l^2}
           +{1\over 2\k}(b_4-b_6)\ul R
           =-{1\over \k}(b_4+b_6)R_{\bot \bot},\label{scm} \\
  {\anti \phi}_{\ol{\mu \nu}\bot}\equiv &&{{\anti \pi}
           _{\ol{\mu \nu}\bot}\over J}+{1\over \k}
           (b_2-b_5){\ul R}_{[\ol{\mu \nu}]}
           =-{1\over \k}(b_2+b_5) R_{\bot [\ol
           {\mu \nu}]\bot},\\
  {\symm \phi}_{\ol{\mu \nu}\bot}\equiv &&{{\symm \pi}
           _{\ol{\mu \nu}\bot}\over J}+{1\over \k}
           (b_1-b_4){\ul R}_{\lan \ol{\mu \nu}\ran}
           =-{1\over \k }(b_1+b_4)R_{\bot \lan \ol
           {\mu \nu}\ran \bot},
 \end{eqnarray}
 \setcounter{equation}{\value{enumi}}}
where ${\ps R}_{\circ \bot}:=\e ^{\mu \nu \s \bot}R_{\ol {\mu \nu
\s}\bot}$,
${\ps R}_{\bot \circ}:=\e ^{\mu \nu \s \bot}R_{\bot \ol {\mu \nu
\s}}$,  $\ul R_{\ol{\mu \nu}}:=R_{\ol{\s \mu \nu}}{}^{\ol \s}$, and
$\ul R:=\ul R_{\ol \mu}{}^{\ol \mu}$.  By a similar argument as used
above, for various degenerate parameter combinations
 one can obtain any of the six expressions of (\ref{323a}-f) as PIC's.
The relations between the critical parameter combinations and the
constraints are summarized in table 1.

 In order to treat all such possibilities in a concise way,
 the singular function
 \begin{equation}
  {\la (x)\over x}\equiv \left\{ \begin{array}{ll}
    1/x, & x\ne 0, \\
    0, &x=0,\end{array} \right.
 \end{equation}
    was introduced.
  The PIC's in the total Hamiltonian density (\ref{thd}) can be given
  in the form
  \begin{equation}
  u^A\phi _A=(u\cdot \phi)^T+(u\cdot \phi)^R,
  \end{equation}
  where
{\setcounter{enumi}{\value{equation}}
 \setcounter{equation}{0}
 \renewcommand{\theequation}{\arabic{section}.\theenumi\alph{equation}}
  \begin{eqnarray}
   (u\cdot \phi)^T\equiv &&[1-\la (2a_1+a_2)]u^{\ol \mu \bot}
    \phi _{\ol \mu \bot}+[1-\la (a_1)]{\symm u}{}^{\ol{\mu \nu}}
    {\symm \phi}_{\ol{\mu \nu}}\nonumber \\
    &&+[1-\la (a_1+2a_3)]{\anti u}{}^{\ol{\mu \nu}}
    {\anti \phi}_{\ol{\mu \nu}}+{1\over 3}[1-\la (a_2)]u\phi
  \end{eqnarray}
  and
  \begin{eqnarray}
   (u\cdot \phi)^R\equiv &&{1\over 6}[1-\la (b_2+b_3)]\ps u\ps \phi
   +{4\over 3}[1-\la (b_1+b_2)]{\tens u}{}^{\ol {\s \mu \nu}}
   {\tens \phi}_{\ol{\s \mu \nu}}\nonumber \\
   &&+[1-\la (b_4+b_5)]{\vect u}{}^{\ol \mu}{\vect \phi}_{\ol \mu}
   +2[1-\la (b_2+b_5)]{\anti u}{}^{\ol{\mu \nu}\bot}
          {\anti \phi}_{\ol{\mu \nu}\bot}\nonumber \\
   &&+2[1-\la (b_1+b_4)]{\symm u}{}^{\ol{\mu \nu}\bot}{\symm \phi}
   _{\ol{\mu \nu}\bot}+{2\over 3}[1-\la (b_4+b_6)]u^\bot\phi _\bot.
  \end{eqnarray}
 \setcounter{equation}{\value{enumi}}}
 The super-Hamiltonian $\h _\bot$ in $\h _{\hbox{\footnotesize can}}$
 (\ref{chd}) then turns out to be of the form
 \begin{equation}
  \h _\bot=\h ^T_\bot +\h ^R_\bot,
 \end{equation}
 with
{\setcounter{enumi}{\value{equation}}
 \setcounter{equation}{0}
 \renewcommand{\theequation}{\arabic{section}.\theenumi\alph{equation}}
 \begin{eqnarray}
  \h ^T_\bot=&&-{1\over 2}Jl^2\left[ {3\la (2a_1+a_2)\over 2a_1+a_2}
  \phi _{\ol \mu \bot}\phi ^{\ol \mu \bot}
  +{3\la (a_1+2a_3)\over a_1+2a_3}{\anti \phi}_{\ol{\mu \nu}}
  {\anti \phi}{}^{\ol{\mu \nu}}\right.\nonumber \\
  &&\left.\quad +{\la (a_1)\over a_1}{\symm \phi}_{\ol{\mu \nu}}
  {\symm \phi}{}^{\ol{\mu \nu}}
  +{\la (a_2)\over 3a_2}\phi ^2\right] -
    J{\ul L}^T-n^\mu\na _a\pi ^a{}_\mu,\\
  \h ^R_\bot=&&-J\k\left[ {\la (b_2+b_3)\over 24(b_2+b_3)}{\ps \phi}^2
   +{\la (b_4+b_5)\over 4(b_4+b_5)}{\vect \phi}_{\ol \mu}
   {\vect \phi}{}^{\ol \mu}\right.\nonumber \\
   &&\quad
   +{\la (b_1+b_2)\over 3(b_1+b_2)}{\tens \phi}_{\ol{\s \mu \nu}}
   {\tens \phi}{}^{\ol{\s \mu \nu}}
  +{\la (b_2+b_5)\over 2(b_2+b_5)}{\anti \phi}_{\ol{\mu \nu}\bot}
  {\anti \phi}{}^{\ol{\mu \nu}\bot}\nonumber \\
  &&\left.\quad
  +{\la (b_1+b_4)\over 2(b_1+b_4)}{\symm \phi}_{\ol{\mu \nu}\bot}
  {\symm \phi}{}^{\ol{\mu \nu}\bot}
  +{\la (b_4+b_6)\over 6(b_4+b_6)}\phi _\bot \phi ^\bot\right] -
    J{\ul L}^R,
 \end{eqnarray}
 where
 \begin{eqnarray}
  {\ul L}^T=&&{1\over 12l^2}\left[ (2a_1+a_3)T_{\ol{\nu \s}\mu}
  T^{\ol{\nu \s}\mu}\right.\nonumber \\
  &&\left.\qquad +2(a_1-a_3)T_{\ol{\nu \s \mu}}T^{\ol{\mu \s \nu}}
  -2(a_1-a_2){\vect T}_{\ol \mu}{\vect T}{}^{\ol \mu}\right] ,\\
  {\ul L}^R=&&-c_0R_{\ol{\tau \s}\mu \nu}R^{\ol{\tau \s}\mu \nu}
  -c_1R_{\ol{\tau \s \mu}\nu}R^{\ol{\tau \mu \s}\nu}\nonumber \\
  &&\quad -c_2R_{\ol{\tau \s \mu \nu}}R^{\ol{\mu \nu \tau \s}}
  -c_3({\ul R}_{\ol{\mu \nu}}{\ul R}^{\ol{\mu \nu}}+R_{\ol \mu \bot}
  R^{\ol \mu \bot})\nonumber \\
  &&\qquad -c_4{\ul R}_{\ol{\mu \nu}}{\ul R}^{\ol{\nu \mu}}
  -c_5{\ul R}^2-{a_0\over 2l^2}{\ul R}+\La,
 \end{eqnarray}
 \setcounter{equation}{\value{enumi}}}
 $\La$ is the cosmological constant.
  The relations between the constants $c_i$'s and the $b_i$'s are
  given by
{\setcounter{enumi}{\value{equation}}
 \addtocounter{enumi}{1}
 \setcounter{equation}{0}
 \renewcommand{\theequation}{\arabic{section}.\theenumi\alph{equation}}
  \begin{eqnarray}
  c_0=&&-{1\over 24\k}(2b_1+3b_2+b_3),\\
  c_1=&&-{1\over 6\k}(b_1-b_3),\\
  c_2=&&-{1\over 24\k}(2b_1-3b_2+b_3),\\
  c_3=&&{1\over 4\k}(b_1+b_2-b_4-b_5),\\
  c_4=&&{1\over 4\k}(b_1-b_2-b_4+b_5),\\
  c_5=&&-{1\over 24\k}(2b_1-3b_4+b_6).
 \end{eqnarray}
 \setcounter{equation}{\value{enumi}}}
 We will now apply this wonderful general ``if-constraint''
Hamiltonian
formulation  to certain spin-zero modes of PGT with specific
parameter combinations.


\section{Spin-zero modes}
\setcounter{equation}{0}
The PGT propagating modes with a single spin-zero propagating mode are
supposed to have ghost-free and tachyon-free
characters\cite{Hayash80,Sezgin80a,Sezgin80b,Battit85,Hecht96}.  There are two
spin-zero modes, i.e., $0^+$ and $0^-$ according to table 1.  These
spin-zero modes were analyzed by a covariant Lagrangian technique and
were shown to have a well posed initial value problem with no tachyonic
propagation characteristics\cite{Hecht96}.  In order to better
understand them, and the non-linear PGT problems, it is
worthwhile to examine their full nonlinear behavior under the
Hamiltonian analysis and how (and indeed whether) they really avoid
the ``constraint bifurcation'' problems (which could be expected to
arise from the many nonlinear constraints).

The restrictions $a_1+2a_3=0$ and $2a_1+a_2=0$ have been regarded as
``viable'' conditions for the PGT theories, we will likewise (and in
the interests of simplicity) also assume these restrictions in both of
our propagating spin-zero parameter choices.


\subsection{Spin-$0^+$ mode}

According to Table 1.
$\pi _\bot$ corresponds to spin-$0^+$.  We
make the specific parameter choices:
 \begin{eqnarray}
   &&2a_1+a_2=0,\quad a_1+2a_3=0,\nonumber \\
   &&b_1=b_2=b_3=b_4=b_5=0,\label{0ppc} \\
   &&a_0\ne 0,\quad b_6\ne 0.\nonumber
 \end{eqnarray}
In fact it is not necessary to take
$b_1$, $\dots$, $b_5$  to vanish.  The
choice can be relaxed by requiring only $b_1=-b_2=b_3=-b_4=b_5\ne 0$
(refer to (\ref{323a}-f)).  However, we believe that it's more critical to
understand the character of the pure spin-$0^+$ mode.  This simple
choice will greatly simplify calculations, while more general choices
are expected to have the same qualitative behavior.   After all
the 4-covariant
analysis\cite{Hecht96}, has argued that there is a 4-parameter class of
Lagrangians with a dynamic spin-$0^+$, but each has qualitatively the same
dynamic behavior.  Hence, we expect that the Hamiltonian analysis of these
more general choices, aside from being calculationally much more
complicated and less transparent, would actually show no new
interesting dynamic features, so we leave them for future work.

 The corresponding super-Hamiltonian is:
 \begin{equation}
  \h _\bot^+=\h ^T_\bot+\h ^{R+}_\bot,
 \end{equation}
{\setcounter{enumi}{\value{equation}}
 \setcounter{equation}{0}
 \renewcommand{\theequation}{\arabic{section}.\theenumi\alph{equation}}
 and
 \begin{eqnarray}
 \h ^T_\bot=&&-{l^2\over 2Ja_1}{\symm \pi}_{\ol {\mu \nu}}
            {\symm \pi}{}^{\ol {\mu \nu}}
            +{l^2\over 12Ja_1}\pi ^2
            -n^\mu \na _c\pi ^c{}_\mu -J{\ul L}^T,\label{htp}\\
 \h ^{R+}_\bot=&&{J\k \over 6b_6}({\pi _\bot \over J}
               +{3a_0\over l^2}-{b_6\over 2\k}{\ul R})^2-
               {Jb_6\over 24\k}{\ul R}^2+{a_0\over 2l^2}J{\ul R}-J\La,
               \label{hcpp}\\
 {\ol L}^T=&&{a_1\over 8l^2}T_{\ol {\nu \s}
             \mu }T^{\ol {\nu \s}\mu }+{a_1\over 4l^2}
             T_{\ol {\nu \s \mu}}T^{\ol {\mu \s \nu}}-
             {a_1\over 2l^2}{\vect T}_{\ol \mu}
             {\vect T}{}^{\ol \mu}.
 \end{eqnarray}
 \setcounter{equation}{\value{enumi}}}

 Due to the parameter choice (\ref{0ppc}), the PIC's are as follows:
 \begin{eqnarray}
  \phi _{\ol \mu \bot }
  \equiv &&{\pi _{\ol \mu \bot }\over J}-
           {a_1\over l^2}{\vect T}_{\ol \mu}\approx 0,\\
  {\anti \phi }_{\ol {\mu \nu }}\equiv &&{{\anti \pi }
           _{\ol {\mu \nu }}\over J}-{a_1\over 2l^2}T_{
           \ol {\mu \nu }\bot }\approx 0,\\
  \ps \phi \equiv &&{\ps \pi \over J}\approx 0,\\
  {\vect \phi}_{\ol \mu}\equiv &&{{\vect \pi}_{\ol \mu}\over J}\approx 0,\\
  {\tens \phi}_{\ol{\s \mu \nu}}\equiv &&{{\tens \pi}
           _{\ol{\s \mu \nu}}\over J}\approx 0,\\
  {\anti \phi}_{\ol{\mu \nu}\bot}\equiv &&{{\anti \pi}
           _{\ol{\mu \nu}\bot}\over J}\approx 0,\\
  {\symm \phi}_{\ol{\mu \nu}\bot}\equiv &&{{\symm \pi}
           _{\ol{\mu \nu}\bot}\over J}\approx 0.
 \end{eqnarray}
  According to Dirac-Bergmann algorithm it's necessary to identify
  the class of these constraints. The
  non-vanishing Poisson brackets (PB) for the PIC's are the following:
\begin{eqnarray}
   \{ {\anti \phi}_{\ol {\mu \nu }\bot}, {\anti \phi}{}_{\ol {\tau
   \s}}^\prime \} \approx &&{\de _{xx^\prime }\over 3J}\Dep
   \eta _{\ol \s [\ol \mu}\eta _{\ol \nu ]\ol \tau},\\
   \{ {\vect \phi}_{\ol \mu }, \phi _{\ol \nu \bot }^\prime \} \approx
   &&-{2\de _{xx^\prime }\over 3J}\Dep
   \eta _{\ol {\mu \nu}}.
\end{eqnarray}
 where ${\displaystyle {\Dep :={\pi _\bot \over J}+{3a_1\over l^2}}}$.
 Thus $\phi _{\ol \mu \bot}$, $\anti \phi _{\ol {\mu \nu}}$,
 $\vect \phi _{\ol \mu}$ and $\anti \phi _{\ol {\mu \nu}\bot}$ are
 {\it second-class} --- as long as $\Dep \ne 0$. The constraints $\ps
\phi$,
 $\tens \phi _{\ol {\s \mu \nu}}$ and $\symm \phi _{\ol {\mu \nu}
 \bot}$ commute with the other primary constraints; according to the
general PGT ``if'' constraint analysis\cite{Blagoj83,Nikoli84}, the
associated SC constraints can
be derived from them  by calculating their time derivatives. They are
 \begin{eqnarray}
   \ps {\dot \phi}\approx &&
   -{1\over 3}N\Dep \ps T\approx 0,\\
   \dot {\symm \phi }_{\ol {\mu \nu }\bot }
   \approx &&-{l^2\over 3a_1}N\Dep
   {{{\symm \pi }_{\ol {\mu \nu }}}\over J}\approx 0,\\
   \tens {\dot \phi}_{\ol {\s \mu \nu }}
   \approx &&-{1\over 3}N\Dep
   {\tens T}_{\ol {\mu \nu \s }}\approx 0.
 \end{eqnarray}
For the time being we put aside the highly degenerate case
 $\Dep \approx 0$.  Then, assuming that $\Dep$ does not vanish
(except perhaps on a set of measure zero),
these
SC's can be  simplified to  \begin{eqnarray}
 {\ps \chi}\equiv &&\ps T\approx 0,\\
 {\symm \chi}_{\ol {\mu \nu}\bot} \equiv &&{{{\symm \pi }
 _{\ol {\mu \nu }}}\over J}\approx 0,\\
 {\tens \chi}_{\ol {\s \mu \nu}}\equiv &&\tens T_{\ol {\mu \nu \s}}
 \approx 0.
 \end{eqnarray}
 By identifying and
eliminating all primary and secondary constraints inside, the
supermomenta and Lorentz rotation parts lead to the distinct
constraints:
 \begin{eqnarray}
 \h _a^+\approx 0\quad \Rightarrow &&-{\pi _\bot \over J}R_{a\bot}
 -\p _a {\pi \over J}+{3a_1\over l^2}{\vect T}{}^bn_\mu \na _b
 \th _a{}^\mu \approx 0,\\
 \h _{\mu \nu}^+\approx 0 \quad \Rightarrow && \Dep T_{\ol {\mu
\nu}\bot}  \approx 0, \\
 &&\Dep \vect T_{\ol \mu}+e^a{}_{\ol \mu}\p _a\Dep \approx 0.\label{cbp}
 \end{eqnarray}

 All of the ``if'' constraints (including the PIC's and the SC's)
are eventually second-class by calculating their PB's with the others.
The details are shown in the appendix.
 The consistency conditions have to be obeyed,
 i.e., we must produce the time derivatives of second-class
 constraints and force the results to vanish weakly.  We then learn the
 Lagrange multipliers. The procedure can be
understood  from the following expression:
 \begin{eqnarray}
  {\dot \phi _B} \equiv &&\int \{ \phi _B, H^\prime
  _{\hbox{\footnotesize tot}}\} \nonumber \\
   =&&\int [\{ \phi _B, \h _{\hbox{\footnotesize can}}^\prime \}
   +u^{\prime A}\{\phi _B, \phi^\prime _A\}]\approx 0.\label{sp}
 \end{eqnarray}
 If $\phi _B$ is second-class, then it is possible to solve for some $u^A$.
 In $\h _{\hbox{\footnotesize tot}}$ (\ref{thd}) only the super-Hamiltonian
 $\h _\bot$ (namely $\h ^+_\bot$ here)
 involving time evolution is important.  By proceeding with the step
described by (\ref{sp}), it is straightforward to get the Lagrange
multipliers:
{\setcounter{enumi}{\value{equation}}
 \addtocounter{enumi}{1}
 \setcounter{equation}{0}
 \renewcommand{\theequation}{\arabic{section}.\theenumi\alph{equation}}
 \begin{eqnarray}
  u^{\ol \mu \bot}=&&-{1\over 2}NJ{\vect T}{}^{\ol \mu},\quad
  {\anti u}{}^{\ol {\mu \nu}}=0,\\
  {\vect u}{}^{\ol \mu}=&&{1\over 2}NJR^{\ol \mu \bot},\quad
  {\anti u}{}^{\ol {\mu \nu}\bot}=-{1\over 2}NJ{\ul R}^{[\ol {\mu \nu}]},\\
  {\ps u}=&&{1\over 4}NJ{\ps R}_{\circ \bot},\\
  {\tens u}{}^{\ol {\s \mu \nu}}=&&{1\over 2}NJ\tens R
    ^{\ol {\mu \nu \s}\bot},\\
  {\symm u}{}^{\ol {\mu \nu}\bot}
    =&&{1\over 2}NJ\left[ {\pi \over \pi _\bot}\eta ^{\ol \s \lan \ol \mu |}
    e^{a|\ol \nu \ran }\na _an_\s
   -{3a_1\over l^2}{J\over \pi _\bot}
    \eta ^{\ol \s \lan \ol \mu |}e^{a|\ol \nu \ran }\na _a
 {\vect T}_{\ol \s}-{\ul R}^{\lan \ol {\mu \nu}\ran}\right] .
 \end{eqnarray}
 \setcounter{equation}{\value{enumi}}}

Now the canonical Hamilton equations of motion can be derived
directly from the completed total Hamiltonian density,
 \begin{eqnarray}
   \dot q_A=&&\int \{ q_A,\h ^\prime
              _{\hbox{\footnotesize tot}}\},\label{tdcv} \\
   \dot \pi _A=&&\int \{ \pi _A,\h ^\prime
              {\hbox{\footnotesize tot}}\} ,
 \end{eqnarray}
where $q_A$ represents the collection of canonical variable $\th
_i{}^\mu$ and $\G _i{}^{\mu \nu}$, $\pi _A$ represents the collection
of the conjugate momenta $\pi ^i{}_\mu$ and $\pi ^i{}_{\mu \nu}$.


\subsection{Massive spin-$0^-$ mode}

 From table 1, $\ps \pi$
 corresponds to the spin-$0^-$ dynamic mode.  We consider the simple
parameter choice:

 \begin{eqnarray}
   &&2a_1+a_2=0,\quad a_1+2a_3=0,\nonumber \\
   &&b_1=b_2=b_4=b_5=b_6=0,\label{0mpc} \\
   &&a_0\ne 0,\quad b_3\ne 0.\nonumber
 \end{eqnarray}
 The corresponding super-Hamiltonian is:
 \begin{eqnarray}
  \h _\bot^-=&\h ^T_\bot+\h ^{R-}_\bot,
 \end{eqnarray}
 and
{\setcounter{enumi}{\value{equation}}
 \setcounter{equation}{0}
 \renewcommand{\theequation}{\arabic{section}.\theenumi\alph{equation}}
 \begin{equation}
 \h ^{R-}_\bot=-{J\k \over 24b_3}({\ps \pi \over J}
               -{b_3\over \k}\ps R_{\circ \bot})^2-{Jb_3\over 24\k}
               \ps R_{\circ \bot}\ps R_\circ {}^\bot
               +{a_0\over 2l^2}J{\ul R}-J\La.\label{hcpn}
 \end{equation}
 \setcounter{equation}{\value{enumi}}}

 In this case the PIC's are:
 \begin{eqnarray}
  \phi _{\ol \mu \bot }
  \equiv &&{\pi _{\ol \mu \bot }\over J}-
           {a_1\over l^2}{\vect T}_{\ol \mu}\approx 0,\label{427}\\
  {\anti \phi }_{\ol {\mu \nu }}\equiv &&{{\anti \pi }
           _{\ol {\mu \nu }}\over J}-{a_1\over 2l^2}T_{
           \ol {\mu \nu }\bot }\approx 0,\\
  {\vect \phi}_{\ol \mu}\equiv &&{{\vect \pi}_{\ol \mu}\over J}\approx 0,\\
  {\tens \phi}_{\ol{\s \mu \nu}}\equiv &&{{\tens \pi}
           _{\ol{\s \mu \nu}}\over J}\approx 0,\\
  \phi _\bot \equiv &&{\pi _\bot \over J}+{3a_0\over l^2}\approx 0,\\
  {\anti \phi}_{\ol{\mu \nu}\bot}\equiv &&{{\anti \pi}
           _{\ol{\mu \nu}\bot}\over J}\approx 0,\\
  {\symm \phi}_{\ol{\mu \nu}\bot}\equiv &&{{\symm \pi}
           _{\ol{\mu \nu}\bot}\over J}\approx 0.\label{433}
 \end{eqnarray}
  The non-zero PB's for the PIC's are the following:
\begin{eqnarray}
   \{ {\anti \phi}_{\ol {\mu \nu }\bot}, \phi _{\ol \s \bot}
   ^\prime \} \approx &&-{\de _{xx^\prime }\over 6J}
   {\ps \pi \over J}\e _{\mu \nu \s \bot},\label{434}\\
   \{ {\anti \phi}_{\ol {\mu \nu }\bot}, {\anti \phi}{}_{\ol {\tau
   \s}}^\prime \} \approx &&-{\de _{xx^\prime }\over J}m
   \eta _{\ol \s [\ol \mu}\eta _{\ol \nu ]\ol \tau},\\
   \{ {\vect \phi}_{\ol \mu}, \phi _{\ol \nu \bot }^\prime \} \approx
   &&2{\de _{xx^\prime }\over J}m
   \eta _{\ol {\mu \nu}},\\
   \{ \vect \phi _{\ol \mu}, {\anti \phi}{}_{\ol {\nu \s}}
   ^\prime \} \approx &&{\de _{xx^\prime }\over 6J}
   {\ps \pi \over J}\e _{\mu \nu \s \bot}\label{437},
\end{eqnarray}
where $\displaystyle {m={a_0-a_1\over l^2}}$.
We first consider the generic ``massive'' case:  $m\ne 0$.
Since the determinant of the PB's matrix is
 $J^{-12}[({\ps \pi}/6J)^2 +m^2]^6>0$, the constraints
 $\phi _{\ol \mu \bot}$, $\anti \phi _{\ol {\mu \nu}}$,
 $\vect \phi _{\ol \mu}$ and $\anti \phi _{\ol {\mu \nu}\bot}$ are
  second-class.

 Similarly we have the time derivatives of $\phi
 _\bot$, ${\symm \phi}_{\ol {\mu \nu }\bot}$ and
 $\tens \phi _{\ol {\s \mu \nu }}$,
  \begin{eqnarray}
 {\dot \phi}_\bot \approx &&{1\over 12}{\ps T}{\ps \pi \over J}+m{l^2\over a_1}
 {\pi \over J}\approx 0\label{438}\\
 \dot {\symm \phi }_{\ol {\mu \nu }\bot}
 \approx &&{1\over 9}{\ps \pi \over J}
 \e _{(\mu |}{}^{\tau \s \bot}\tens T_{\ol {\tau \s}|\ol \nu)}+
 m{l^2\over a_1}{{{\symm \pi }
 _{\ol {\mu \nu }}}\over J}\approx 0,\label{439}\\
 \tens {\dot \phi}_{\ol {\s \mu \nu }}
 \approx &&{l^2\over 6a_1}{\ps \pi \over J^2}
 \e _{\tau \mu (\nu |\bot}{\symm \pi}_{|\ol \s )}{}^{\ol \tau}+m
 \tens T_{\ol {\mu \nu \s}}
 \approx 0.\label{440}
 \end{eqnarray}
Generically these can be replaced by the three simpler SC
constraints:
  \begin{eqnarray}
 \chi _\bot \equiv &&{1\over 12}{\ps T}{\ps \pi \over J}+m{l^2\over a_1}
 {\pi \over J}\approx 0,\\
 \symm \chi _{\ol {\mu \nu}\bot} \equiv
 &&{{{\symm \pi }
 _{\ol {\mu \nu }}}\over J}\approx 0,\\
 \tens \chi _{\ol {\s \mu \nu}}\equiv
 &&\tens T_{\ol {\mu \nu \s}}
 \approx 0.
 \end{eqnarray}
 The supermomenta and Lorentz rotation parts are:
 \begin{eqnarray}
 \h _a^-\approx 0\Rightarrow &&
 {a_1\over 4l^2}{\vect T}{}^{\ol \nu}
 T_{\ol {\mu \nu}\bot}
 +{a_1\over 12l^2}\ps T\e _{\mu \nu \s \bot}T^{\ol {\nu \s}}{}_\bot
 \nonumber \\
 &&\quad -{1\over 12}{\ps \pi \over J}{\ps R}_{\ol \mu \circ}
 -{a_0\over l^2}R_{\ol \mu \bot}
 -{a_1\over l^2}{\vect T}{}^a
 \na _an_\mu \nonumber \\
 &&\quad -{a_1\over 2l^2}\eta _{\ol {\mu \nu}}e^a{}_{\ol \s}\na _a
 T^{\ol {\nu \s}}{}_\bot+{1\over 3}e^a{}_{\ol \mu}\p _a
 {\pi \over J}\approx 0,\\
 \h _{\mu \nu}^-\approx 0 \Rightarrow &&{1\over 12}{\ps \pi \over J}
 \e _{\mu \nu \s \bot}T^{\ol {\nu \s}}{}_\bot -m\vect T_{\ol \mu}
 \approx 0,\\
 &&{1\over 6}\e _{\mu \nu \s \bot}(e^{a\ol \s}\p _a{\ps \pi \over J}
 +{\ps \pi \over J}{\vect T}{}^{\ol \s})+mT_{\ol {\mu \nu}\bot}
 \approx 0,\label{ncbp}
 \end{eqnarray}
 where
 ${\ps R}_{\ol \mu \circ}:=\e ^{\nu \tau \s \bot}R_{\ol {\mu \nu
 \tau \s}}$.
 Utilizing again the standard Dirac-Bergmann algorithm the corresponding
 Lagrange multipliers are found to be
{\setcounter{enumi}{\value{equation}}
 \addtocounter{enumi}{1}
 \setcounter{equation}{0}
 \renewcommand{\theequation}{\arabic{section}.\theenumi\alph{equation}}
 \begin{eqnarray}
 u^{\ol \mu \bot}=&&-{1\over 2}NJ{\vect T}{}^{\ol \mu},\quad
 {\anti u}{}^{\ol {\mu \nu}}=NJT^{\ol {\mu \nu}}{}_\bot ,\\
 {\vect u}{}^{\ol \mu}=&&{1\over 2}NJR^{\ol \mu \bot},\quad
 {\anti u}{}^{\ol {\mu \nu}\bot}=-{1\over 2}NJ{\ul R}^{[\ol {\mu \nu}]},\\
 u^\bot =&&-{a_1\over 4a_0}NJ\left[ {3\over 8}T^{\ol {\mu \nu}}
 {}_\bot T_{\ol {\mu \nu}\bot}+{3\over 4}{\vect T}_{\ol \mu}
 {\vect T}{}^{\ol \mu}-{1\over 16}{\ps T}^2+{a_0\over 2a_1}{\ul R}
 -{2l^2\over a_1}\La\right.
 \nonumber \\
 &&\left. -{l^4\over 4a^2_1}({\pi \over J})^2+{\k \over b_3}
 ({1\over 6m}+{l^2\over a_1})({\ps \pi \over J})^2
 -({1\over 12m}+{2l^2\over a_1}){\ps R}_{\circ \bot}{\ps \pi \over J}
 \right] ,\\
 {\tens u}{}^{\ol {\s \mu \nu}}
 =&&{1\over 4}NJ\left[ \tens (T^{\ol {\mu \nu}\bot}{\vect T}{}^{\ol \s})
 +{1\over 3}{\ps T}\tens (\e ^{\mu \nu \tau \bot}e^{a\ol \s})\na _an_\tau
 +2\tens R^{\ol {\mu \nu \s}\bot}\right.\nonumber \\
 &&\left. -2\tens (\eta ^{\ol {\mu \tau}}\eta ^{\ol {\nu \vph}}e^{a\ol \s})
 \na _aT_{\ol {\tau \vph}\bot}\right] ,\\
 {\symm u}{}^{\ol {\mu \nu}\bot}
 =&&{1\over 2}NJ\left[
 -{a_1\over 2a_0}T^{\lan \ol \mu |\ol \s \bot}T^{|\ol \nu \ran}
 {}_{\ol \s \bot}
 +{l^2\over 3a_0}m{\vect T}{}^{\lan \ol \mu}{\vect T}{}^{\ol \nu \ran}
 +{l^2\over 36a_0}{\ps \pi \over J}\e ^{\s \tau \lan \mu |\bot}
 T^{|\ol \nu \ran}{}_{\ol \s \bot}\vect T_{\ol \tau}\right.\nonumber \\
 &&-{l^2\over 3a_1}{\pi \over J}
 \eta ^{\ol \s \lan \ol \mu |}e^{a|\ol \nu \ran}
  \na _an_\s +{a_1\over a_0}
 \eta ^{\ol \s \lan \ol \mu |}e^{a|\ol \nu \ran }\na _a
 {\vect T}_{\ol \s}-{\ul R}{}^{\lan \ol {\mu \nu}\ran}\nonumber \\
 &&\left. -{l^2\over 9a_0}{\ps \pi \over J}\tens (\e ^{\lan \mu |\tau \s \bot}
 e^{a|\ol \nu \ran})\na _aT_{\ol {\tau \s}\bot}\right] .
 \end{eqnarray}
 \setcounter{equation}{\value{enumi}}}


\subsection{Massless spin-$0^-$ mode}
 In the spin-$0^-$ mode there is a special ``massless'' subcase
given by
 \begin{eqnarray*}
m=0 ,\quad \mbox{i.e.,} \quad a_1=a_0.
 \end{eqnarray*}
 We take the other parameters in the massless
case to be the same
 as those of the massive spin-$0^-$ mode.
Therefore we will only mention the parts that differs from those
 in the massive spin-$0^-$ case. The supermomenta and Lorentz rotation parts can
 be further simplified
 \begin{eqnarray}
 \h _a^-\approx 0\Rightarrow &&
 -{1\over 12}{\ps \pi \over J}{\ps R}_{\ol \mu \circ}
 -{a_0\over l^2}R_{\ol \mu \bot}\nonumber \\
 &&\quad -{a_0\over l^2}{\vect T}{}^a \na _an_\mu
 +{1\over 3}e^a{}_{\ol \mu}\p _a
 {\pi \over J}\approx 0,\\
 \h _{\mu \nu}^-\approx 0 \Rightarrow &&
 {\ps \pi} T_{\ol {\mu \nu}\bot}
 \approx 0,\label{Hmunu-a}\\
 &&\p _a{\ps \pi \over J}
 +{\ps \pi \over J}{\vect T}_a
 \approx 0.
 \end{eqnarray}
 The non-zero PB's for the PIC's that remain, eqs (\ref{434}, \ref{437}), are
 $\{ {\anti \phi}_{\ol {\mu \nu }\bot}, \phi _{\ol \s \bot}
   ^\prime \}$,
  $\{ \vect \phi _{\ol \mu}, {\anti \phi}{}_{\ol {\nu \s}}
  ^\prime \}$. Whether the PB's matrix is singular is determined
 by ${\ps \pi}/J$.
Let us first suppose the ${\ps \pi}$ does not vanish (except perhaps
on a set of measure zero).
Then the class of
 $\phi _{\ol \mu \bot}$, ${\anti \phi}_{\ol {\mu \nu}}$,
 ${\vect \phi}_{\ol \mu}$, and ${\anti \phi}_{\ol{\mu \nu }\bot}$
won't change. The SC's are
 \begin{eqnarray}
 \chi _\bot \equiv &&\ps T
 \approx 0\\
 \symm \chi _{\ol {\mu \nu}\bot} \equiv
 &&{{{\symm \pi }
 _{\ol {\mu \nu }}}\over J}\approx 0,\\
 \tens \chi _{\ol {\s \mu \nu}}\equiv
 &&\tens T_{\ol {\mu \nu \s}}
 \approx 0.
 \end{eqnarray}
 The Lagrange multipliers are
{\setcounter{enumi}{\value{equation}}
 \addtocounter{enumi}{1}
 \setcounter{equation}{0}
 \renewcommand{\theequation}{\arabic{section}.\theenumi\alph{equation}}
 \begin{eqnarray}
 u^{\ol \mu \bot}=&&-{1\over 2}NJ{\vect T}{}^{\ol \mu},\quad
 {\anti u}{}^{\ol {\mu \nu}}=0,\label{454a}\\
 {\vect u}{}^{\ol \mu}=&&{1\over 2}NJR^{\ol \mu \bot},\quad
 {\anti u}{}^{\ol {\mu \nu}\bot}=-{1\over 2}NJ{\ul R}^{[\ol {\mu \nu}]},\\
 {\tens u}{}^{\ol {\s \mu \nu}} =&&{1\over 2}NJ
  \tens R^{\ol {\mu \nu \s}\bot}\\
 {\symm u}{}^{\ol {\mu \nu}\bot}=&&{1\over 2}NJ\left[
 -{l^2\over 3a_1}{\pi \over J}
 \eta ^{\ol \s \lan \ol \mu |}e^{a|\ol \nu \ran}\na _an_\s +
 \eta ^{\ol \s \lan \ol \mu |}e^{a|\ol \nu \ran }\na _a
 {\vect T}_{\ol \s}-{\ul R}{}^{\lan \ol {\mu \nu}\ran}\right] .
 \end{eqnarray}
 \setcounter{equation}{\value{enumi}}}
 We find that $\chi _\bot$ commutes with $\phi _\bot$, so its
 consistency condition leads to, instead of determining $u^\bot$,
 a tertiary constraint (a quite unusual occurrence):
 \begin{equation}
  \tc _\bot \equiv {\ps \pi \over J}-{b_3\over
  2\k}{\ps R}_{\circ\bot}\approx 0.
 \end{equation}
 However, $\tc_\bot$ still commutes with $\phi_\bot$ thus $u^\bot$ remains
undetermined. Then the consistency condition of $\tc _\bot$, i.e., ${\dot \tc}
 _\bot \approx 0$, and the substitution of the known multipliers in
 (\ref{454a}-d)
lead, in turn to the constraint 
 \begin{equation}
  \fc \equiv {\ps R}_{\circ \bot}\approx 0.
 \end{equation}
(Because of (\ref{Hmunu-a}) the constraint $\fc\approx0$ can be derived
directly from
the Bianchi identity $\na _{[a}T_{bc]}{}^\mu =R_{[abc]}{}^\mu$.)\ \ Then there
is no need to go further, since the constraint $\tc _\bot$, along with
$\fc$, contradicts our assumption that $\ps\pi$ does not vanish.
Consequently, for the masless spin-$0^-$ case, $\ps\pi$ must definitely vanish.

  Knowing
now that $\ps\pi$ definitely vanishes, so then does
the rhs of (\ref{438},\ref{439},\ref{440}), which tells us that the massless
spin-$0^-$ doesn't have any propagating torsion modes and is, consequently,
just equivalent to GR.


\section{Discussion}
\setcounter{equation}{0}
In the PGT, there are forty dynamic variables coming from sixteen
tetrad components and twenty-four connection components.  The number
of total variables counts eighty because the same number of canonical
momenta accompany the variables.  Nonetheless, constraints eliminate
many unphysical variables.  The super-Hamiltonian $\h
_\bot$, super-momenta $\h _a$, and the Lorentz generators $\h_{\mu
\nu}$ are (generally) ten first-class constraints.  There are also ten ``sure''
first-class primary constraints.  This total of twenty first-class
constraints (because of their gauge nature) offset forty variables---in
general.

In the dynamic spin-$0^+$ and massive
spin-$0^-$ modes that we have considered here, we found that the
twenty-three PIC's and eleven
SC's are second-class.  They thus eliminate thirty-four unphysical
variables.  Therefore, there are only $80-40-34=6$ true physical
variables.  It simply means that the number of degrees of freedom is
reduced to three:  the scalar/pseudoscalar mode and the two helicity
states of the usual massless graviton.

However, for the {\it massless} spin-$0^-$
mode, we have shown that the pseudoscalar canonical momentum must
vanish weakly in order to make the theory self-consistent.  In that case we
have the usual $4+4$ ``sure'' first class constraints associated with the
translational gauge freedom and the 6 ``sure'' first class Lorentz gauge
freedom constraints (\ref{scpc}). All of the ``if'' constraints
(\ref{427}-\ref{433}) then
turn out to be first class.  Their consistency conditions, as well as the
consistency condition for (\ref{scpc}), degenerate to the single first class
condition $\ps \pi\approx0$.  Thus we have a total of $4+4+6+23+1=38$ first
class
constraints leading to $80-2\times38=4$ physical initial values for just the
usual
2 degrees of freedom for the graviton. In this special case the propagating
torsion modes are unphysical, being pure gauge.

One can determine the positivity of the non-zero coupling parameters from the
Hamiltonian density (\ref{htp}, \ref{hcpp}, \ref{hcpn}).  Since the
kinetic energy density must be positive definite, we have
 \begin{equation}
  a_1>0,\quad {b_3\over \k}<0,\quad {b_6\over \k}>0.
 \end{equation}
 The results are consistent with restrictions that earlier researchers
 have proposed, e.g., references\cite{Hayash80,Sezgin80a,Sezgin80b,Battit85}.

It is no wonder that the Lagrange multipliers are identical to the
missing velocities.  Let us examine these in the
spin-$0^+$
mode.  The time derivatives of canonical variables are given in
(\ref{tdcv}). After reducing the results we truly find
{\setcounter{enumi}{\value{equation}}
 \addtocounter{enumi}{1}
 \setcounter{equation}{0}
 \renewcommand{\theequation}{\arabic{section}.\theenumi\alph{equation}}
 \begin{eqnarray}
  T_\bot {}^{\ol \mu \bot}\equiv u^{\ol \mu \bot},&\quad &
    T_\bot {}^{[\ol{\mu \nu}]}\equiv {\anti u}{}^{\ol{\mu \nu}},\\
  R_\bot {}^{\ol \mu}\equiv {\vect u}{}^{\ol \mu},&\quad &
    R_\bot {}^{[\ol{\mu \nu}]\bot}\equiv {\anti u}{}^{\ol{\mu \nu}\bot},\\
  {\tens R}_\bot {}^{\ol{\s \mu \nu}}\equiv
    {\tens u}{}^{\ol{\s \mu \nu}},&\quad &
    R_\bot {}^{\lan \ol{\mu \nu}\ran \bot}\equiv
    {\symm u}{}^{\ol{\mu \nu}\bot},\\
  {\ps R}_{\bot \circ}\equiv {\ps u}.&&
 \end{eqnarray}
 \setcounter{equation}{\value{enumi}}}
The situation of the massive spin-$0^-$ mode is much the same.  On the
other
hand, the expressions for the multipliers, the $u$'s, show that each
missing velocity
is determined by non-velocity terms, hence they do not have a
physically independent dynamical status.
This fact reflects the details of the Lagrangian field equations.
Due to the specific parameter combinations (\ref{0ppc})
and (\ref{0mpc}), the coefficients attached to the time derivatives of those
missing velocities vanish and the related field equations become
constraints.

If we look at the multipliers for the two spin-zero modes, $u^{\ol \mu
\bot}$, ${\anti u}{}^{\ol{\mu \nu}}$, ${\vect u}{}^{\ol \mu}$, and
${\anti u}{}^{\ol{\mu \nu}\bot}$ essentially have the same
expressions.  (${\anti u}{}^{\ol{\mu \nu}}$ vanishes in the spin-$0^+$
and massless spin-$0^-$ modes because $T_{\ol{\mu \nu}\bot}\approx 0$
there.) It is clear that, since $\phi _{\ol \mu \bot}$, ${\anti
\phi}_{\ol{\mu \nu}}$, ${\vect \phi}_{\ol \mu}$, and ${\anti
\phi}_{\ol{\mu \nu}\bot}$ contain the variables related to spin-one
modes, they should not be involved in the field equations of spin-zero
modes.  As to the multipliers ${\symm u}{}^ {\ol{\mu \nu}\bot}$ and
${\tens u}{}^{\ol{\s \mu \nu}}$, which are related to the spin-$2^+$
and spin-$2^-$ modes, in both of the modes they are more involved and
complicated because the usual massless spin-two graviton is mixed up
here.

We note that the constraints (\ref{cbp}) in the spin-$0^+$ case deduced
from the Lorentz rotation parts are intriguing.  Since $\Dep$ is
generically a function of spatial coordinates, the constraint will
generally prevent $\Dep$ from vanishing unless it vanishes globally.
However $\Dep$ could vanish permanently and become a new ``constraint''
if its initial value is zero.  (Assuming that ${\vect
T}_{\ol \mu}$ cannot become unbounded.)\ \ From (\ref{scm})
the relation between $\Dep$ and the (affine {\em not Riemannian}) scalar
curvature is given by  \begin{equation}
  \Dep \equiv {b_6\over 2\k}R-3m,\label{53}
 \end{equation}
therefore $\Dep =0$ indicates that the (affine) scalar curvature is equal to
$6\k m/b_6$, a constant. The constant forms a barrier for the scalar
field to cross.  Whether the scalar mode evolves on such an affine
(anti) de-Sitter
spacetime background or is frozen depends upon the
initial conditions.  This forms a new kind of
constraint bifurcation.  The usual type, which
originates from the PB's matrix of the constraints being not of
constant rank, has been linked to acausal propagating
modes\cite{Chen98}.
(Because of the mass terms, as can be seen from
({\ref{ncbp})), the massive spin-$0^-$ mode avoids getting involved in
this new type of constraint bifurcation.)

It is straightforward to work out, from the covariant analysis of the scalar
PGT modes\cite{Hecht96}, the Lagrangian field equations for the
spin-$0^+$,
case with $\Dep=0$.  They just turn out to be those of GR with
the addition of a corrected cosmological constant depending on $\La$ and the
value of the constant (affine) scalar curvature, $6\k m/b_6$.  All coefficients
of the torsion components in the field equations vanish.  Consequently the
torsion is really pure gauge in this case.  A further very special highly
degenerate subcase, with $\Dep=0=m$, needs no additional discussion; it has
just the behavior expected in the $\Dep\to0$ limit.

This covariant analysis is in accord with the corresponding Hamiltonian
analysis which shows that the $4+4$ ``sure'' translational gauge generators,
the $6$ ``sure'' first class rotation gauge generators and the $23$ primary
``if'' constraints give rise to just one  secondary constraint: $\Dep\approx0$.
All the ``if'' constraints turn out to be first class.
Thus we again have a total of $38$ first class constraints and the usual
2 degrees of freedom for the graviton plus purely gauge torsion.

For solutions with special symmetries, of course there are other
possibilities.  In particular, consider the case of homogeneous cosmologies.
If $\Dep$ is dependent on time only, the constraint (\ref{cbp}) simplifies,
thereby allowing for simpler solutions for scalar fields on a
Robertson-Walker-like spacetime.   But here we are concerning ourselves with
the dynamic structure of the general theory, not with the peculiarities of
solutions with special symmetries.




 We remark that the fourth-derivative  gravity which
people have investigated recently has similar degeneracy
problem\cite{Teyssa83}.  Careful
consideration is required to understand if the relation between its
degeneracy and constraint bifurcation phenomenon exits, but such a study
is beyond the purpose of this paper.

\section{Conclusion}
In this paper the Dirac-``if'' constraint Hamiltonian formalism for
the PGT has been
applied to the study of the full nonlinear behavior of its spin-zero
modes with certain specific coupling parameter choices.  It is shown that
by using the Hamiltonian analysis one can clearly identify the degeneracy and
the corresponding constraints, and the true dynamic degrees of freedom
of the PGT as well as the positivity of its coupling parameters.

For the spin-$0^-$ mode, there is a massive, non-ghost pseudoscalar
field, in addition to the usual spin-two graviton, propagating
dynamically.  But the corresponding massless mode has additional
gauge freedom in the torsion modes, yielding a dynamics which is
essentially equivalent to that of GR (with presumably undetectable purely gauge
torsion).
Therefore, the magnitude of the mass $m$ will determine the
viability and detectability of this mode.  If $|m|\gg 0$, we might
expect to find this mode at about the Planck-scale range.  On the
other hand if $|m|\to 0$, the differences between the PGT and GR are
too tiny to be detected.
In both of these situations we
can only obtain almost the same effects as GR on the ordinary scale.  There is
also a scalar field propagating in the spin-$0^+$ mode case.
This mode encounters a type of constraint bifurcation phenomenon, depending on
whether $\Dep$ vanishes, which
divides the phase-space into two subspaces, since $\Dep$ cannot vanish in
general unlesss it vanishes globaly because of the Lorentz rotation
constraint (\ref{cbp}).
However, as
we discussed above, the scalar momentum could be only a function of
time which turns off the effect of the phenomenon.  Then we have a
solution which can be applied to spatially homogeneous cosmological
models.

The constraint bifurcation phenomenon is generally expected to occur
in the full nonlinear PGT with higher spin propagating
modes\cite{Hecht96}.  Somewhat surprisingly, we also see the
``constraint bifurcation'' phenomenon in these spin-zero cases, albeit
in a very simple essentially ``all or nothing'' fashion.

This work is an important complement to the earlier work on PGT scalar
modes\cite{Hecht96} and is a necessary preliminary to our next step:
examining the spin-one and spin-two modes and comparing the nonlinear
results to the linearized ones of PGT.


\section*{Acknowledgments}
This work was supported by the National Science Council of the R.O.C. under
grants No. NSC87-2112-M-008-007 and No. NSC88-2112-M-008-018.

\section*{Appendix: Non-zero Poisson brackets}
 In order to classify the SC constraints and derive the Lagrange
multipliers,  the values of all PB's should be calculated.
 Here the results of the calculations of non-zero PB's
 in  the spin-zero modes of the PGT are presented.
 \subsection*{Results in the spin-$0^+$ mode}
 \setcounter{equation}{0}
 The followings just show that all constraints in the spin-$0^+$ are
second-class provided $\Dep \ne 0$.
{
 \renewcommand{\theequation}{A.\arabic{equation}}
 \begin{eqnarray}
 \{ \ps \chi, {\anti u}{}^{\prime \ol {\mu \nu}}{\anti \phi}{}^\prime
 _{\ol {\mu \nu}} \}\approx &&2\de _{xx^\prime}\e _{\mu \nu}{}^
 {\s \bot}e^a{}_{\ol \s}\na _a{{\anti u}{}^{\ol {\mu \nu}}\over
 J},\\
 \{ \ps \chi, \ps \phi ^\prime
 \}\approx &&24{\de _{xx^\prime}\over J}\\
 \{ \symm \chi _{\ol {\mu \nu}\bot}, u^{\prime \ol \s \bot}
 \phi ^\prime
 _{\ol \s \bot} \}\approx &&\de _{xx^\prime}{a_1\over l^2}
 \eta _{\ol \s \lan \ol \mu}e^a{}_{\ol \nu \ran }
 \na _a{u^{\ol \s \bot}\over J},\\
 \{ \symm \chi _{\ol {\mu \nu}\bot}, {\anti \phi}{}^\prime
 _{\ol {\tau \s}} \}\approx &&{\de _{xx^\prime}\over J}{a_1\over l^2}
 \eta _{\ol \rho [\ol \s}\eta _{\ol \tau ]\lan \ol \mu}e^a{}_{\ol \nu \ran }
 \na _an^\rho ,\\
 \{ \symm \chi _{\ol {\mu \nu}\bot}, {\symm \phi}{}^\prime
 _{\ol {\tau \s}\bot} \}\approx &&-{\de _{xx^\prime}\over 3J}
 {\pi _\bot \over J}
 \eta _{\ol \s \lan \ol \mu}\eta _{\ol \nu \ran \ol \tau},\\
 \{ \tens \chi _{\ol {\s \mu \nu}}, \phi ^\prime _{\ol \tau \bot} \}
 \approx &&{\de _{xx^\prime}\over J}[\tens(\eta _{\ol {\mu \rho}}
 \eta _{\ol {\nu \tau}}e^a{}_{\ol \s})-
 \tens(\eta _{\ol {\mu \tau}}\eta _{\ol {\nu \rho}}e^a{}_{\ol \s})]
 \na _an^\rho ,\\
 \{ \tens \chi _{\ol {\s \mu \nu}}, {\anti u}{}^{\prime \ol {\tau \rho}}
 {\anti \phi}{}^\prime _{\ol {\tau \rho}} \}
 \approx &&\de _{xx^\prime}\tens(\eta _{\ol {\mu \rho}}
 \eta _{\ol {\nu \tau}}e^a{}_{\ol \s})
 \na _a{{\anti u}{}^{\ol {\tau \rho}}\over J},\\
 \{ \tens \chi _{\ol {\s \mu \nu}}, \tens \phi
 ^\prime _{\ol {\tau \rho \vph}} \}
 \approx &&{\de _{xx^\prime}\over J}[\tens(\eta _{\ol \mu (\ol \vph}
 \eta _{\ol \tau )\ol \nu}\eta _{\ol {\s \rho}})-
 \tens(\eta _{\ol {\mu \rho}}\eta _{\ol{\nu \vph}}
 \eta _{\ol{\tau \s}})].
 \end{eqnarray}
 }


 \subsection*{Results in the spin-$0^-$ mode}
 \setcounter{equation}{0}
 {
 \renewcommand{\theequation}{B.\arabic{equation}}
 \begin{eqnarray}
 \{ \chi _\bot, u^{\prime \ol \mu \bot}\phi \prime
 _{\ol \mu \bot} \}\approx &&{\de _{xx^\prime}\over 6J}{\ps \pi \over
 J} \e ^{\mu \nu \s \bot}u_{\ol \mu}{}^\bot T_{\ol {\nu \s}\bot}
 -2\de _{xx^\prime}me^a{}_{\ol \mu}\na _a{u^{\ol \mu \bot}\over J},\\
 \{ \chi _\bot, {\anti u}{}^{\prime \ol {\mu \nu}}{\anti \phi}{}^\prime
 _{\ol {\mu \nu}} \}\approx &&{\de _{xx^\prime}\over 6}{\ps \pi \over
J} \e _{\mu \nu}{}^
 {\s \bot}e^a{}_{\ol \s}\na _a{{\anti u}{}^{\ol {\mu \nu}}\over
 J}+\de _{xx^\prime}{m\over J}{\anti u}{}^{\ol {\mu \nu}}
 T_{\ol {\mu \nu}\bot},\\
 \{ \chi _\bot, \phi ^\prime _\bot
 \}\approx &&-{\de _{xx^\prime}\over J}{6a_0\over a_1}m,\\
 \{ \symm \chi _{\ol {\mu \nu}\bot}, u^{\prime \ol \s \bot}
 \phi ^\prime
 _{\ol \s \bot} \}\approx &&\de _{xx^\prime}{a_1\over l^2}
 \eta _{\ol \s \lan \ol \mu}e^a{}_{\ol \nu \ran }
 \na _a{u^{\ol \s \bot}\over J},\\
 \{ \symm \chi _{\ol {\mu \nu}\bot}, {\anti \phi}{}^\prime
 _{\ol {\tau \s}} \}\approx &&{\de _{xx^\prime}\over J}{a_1\over l^2}
 \eta _{\ol \rho [\ol \s}\eta _{\ol \tau ]\lan \ol \mu}e^a{}_{\ol \nu \ran }
 \na _an^\rho ,\\
 \{ \symm \chi _{\ol {\mu \nu}\bot}, {\symm \phi}{}^\prime
 _{\ol {\tau \s}\bot} \}\approx &&{\de _{xx^\prime}\over J}
 {a_0 \over l^2}
 \eta _{\ol \s \lan \ol \mu}\eta _{\ol \nu \ran \ol \tau},\\
 \{ \tens \chi _{\ol {\s \mu \nu}}, \phi ^\prime _{\ol \tau \bot} \}
 \approx &&{\de _{xx^\prime}\over J}[\tens(\eta _{\ol {\mu \rho}}
 \eta _{\ol {\nu \tau}}e^a{}_{\ol \s})-
 \tens(\eta _{\ol {\mu \tau}}\eta _{\ol {\nu \rho}}e^a{}_{\ol \s})]
 \na _an^\rho ,\\
 \{ \tens \chi _{\ol {\s \mu \nu}}, {\anti u}{}^{\prime \ol {\tau \rho}}
 {\anti \phi}{}^\prime _{\ol {\tau \rho}} \}
 \approx &&\de _{xx^\prime}\tens(\eta _{\ol {\mu \rho}}
 \eta _{\ol {\nu \tau}}e^a{}_{\ol \s})
 \na _a{{\anti u}{}^{\ol {\tau \rho}}\over J},\\
 \{ \tens \chi _{\ol {\s \mu \nu}}, \tens \phi
 ^\prime _{\ol {\tau \rho \vph}} \}
 \approx &&{\de _{xx^\prime}\over J}[\tens(\eta _{\ol \mu (\ol \vph}
 \eta _{\ol \tau )\ol \nu}\eta _{\ol {\s \rho}})-
 \tens(\eta _{\ol {\mu \rho}}\eta _{\ol{\nu \vph}}
 \eta _{\ol{\tau \s}})].
 \end{eqnarray}
 }

\begin{table}[p]
\begin{tabular}{clcl}\hline
 $J^p$ &
 \begin{tabular}{l}
 Kinetic Parameter\\
 Combinations
 \end{tabular}
 & Constraints &
 \begin{tabular}{l}
 Mass Parameter\\
 Combinations
 \end{tabular}
 \\ \hline
 $0^+$ &
    \begin{tabular}{ll}
      (\Ri )&$a_2$\\
      (\Rj )&$b_4+b_6$
    \end{tabular}                 &
    \begin{tabular}{l}
      $\phi$, $\chi$\\
      $\phi _\bot$, $\chi _\bot$
    \end{tabular}                 &
    $a_0$, $2a_0+a_2$
    \\ 
 $1^+$ &
    \begin{tabular}{ll}
    (\Ri )&$a_1+2a_3$\\
    (\Rj )&$b_2+b_5$
    \end{tabular}                 &
    \begin{tabular}{l}
      ${\anti \phi}_{\ol{\mu \nu}}$, ${\anti \chi}_{\ol{\mu \nu}}$\\
      ${\anti \phi}_{\ol{\mu \nu}\bot}$,
      ${\anti \chi}_{\ol{\mu \nu}\bot}$
    \end{tabular}                 &
    $a_1-a_0$, $\displaystyle {{a_0\over 2}+a_3}$
    \\ 
 $2^+$ &
    \begin{tabular}{ll}
    (\Ri )&$a_1$\\
    (\Rj )&$b_1+b_4$
    \end{tabular}                 &
    \begin{tabular}{l}
      ${\symm \phi}_{\ol{\mu \nu}}$, ${\symm \chi}_{\ol{\mu \nu}}$\\
      ${\symm \phi}_{\ol{\mu \nu}\bot}$,
      ${\symm \chi}_{\ol{\mu \nu}\bot}$
    \end{tabular}                 &
    $a_0$, $a_1-a_0$
    \\ 
 $1^-$ &
    \begin{tabular}{ll}
      (\Ri )&$2a_1+a_2$\\
      (\Rj )&$b_4+b_5$
    \end{tabular}                 &
    \begin{tabular}{l}
      $\phi _{\ol \mu \bot}$, $\chi _{\ol \mu \bot}$\\
      ${\vect \phi}_{\ol \mu}$, ${\vect \chi}_{\ol \mu}$
    \end{tabular}                 &
    $a_1-a_0$, $2a_0+a_2$
    \\ 
 $0^-$ &
    \begin{tabular}{ll}
      &$\qquad b_2+b_3$
    \end{tabular}                 &
    \begin{tabular}{l}
      $\ps \phi$, $\ps \chi$
    \end{tabular}&
    $\displaystyle {{a_0\over 2}+a_3}$
\\ 
 $2^-$ &
      \begin{tabular}{ll}
     &$\qquad b_1+b_2$
      \end{tabular}&
      \begin{tabular}{l}
      ${\tens \phi}_{\ol{\s \mu \nu}}$,
      ${\tens \chi}_{\ol{\s \mu \nu}}$
      \end{tabular}&
    $a_1-a_2$
\\ \hline
\end{tabular}
\caption{Primary `if'-constraints, critical parameter values
and masses}
\end{table}

\end{document}